\documentclass[
  aps,
  prb,
  preprint,
  amsmath,
  amssymb,
  superscriptaddress,
  longbibliography
]{revtex4-2}

\usepackage{graphicx}
\usepackage{bm}
\usepackage{xcolor}
\usepackage{booktabs}
\usepackage{hyperref}
\usepackage[acronym,nonumberlist]{glossaries}
\glsdisablehyper
\newacronym{soc}{SOC}{spin--orbit coupling}
\newacronym{dmi}{DMI}{Dzyaloshinskii--Moriya interaction}
\newacronym{nep}{NEP}{neuroevolution potential}
\newacronym{bfo}{BFO}{BiFeO$_3$}
\newacronym{dft}{DFT}{density functional theory}
\newacronym{afm}{AFM}{antiferromagnetic}
\newacronym{snes}{SNES}{separable natural evolution strategy}
\newacronym{rmse}{RMSE}{root-mean-square error}
\newacronym{dzp}{DZP}{double-zeta polarized}
\newacronym{pbe}{PBE}{Perdew--Burke--Ernzerhof}
\newacronym{gpu}{GPU}{graphics processing unit}
\newacronym{md}{MD}{molecular dynamics}
\newacronym{mmtp}{mMTP}{magnetic moment tensor potential}
\newacronym{mf}{MF}{magnetic force}
\newacronym{mae}{MAE}{magnetocrystalline anisotropy energy}
\newacronym{fire}{FIRE}{fast inertial relaxation engine}

\graphicspath{{figures/}}
\newcommand{\vect}[1]{\bm{#1}}
\newcommand{\Rset}{\mathcal{R}}
\newcommand{\Sset}{\mathcal{S}}
\newcommand{\spinvec}{\vect{s}}
\newcommand{\rvec}{\vect{r}}
\newcommand{\mforce}{\vect{M}}
\newcommand{\force}{\vect{F}}
\newcommand{\torque}{\bm{\tau}}
\newcommand{\descriptor}{\vect{q}}
\newcommand{\structdescriptor}[1]{\descriptor^{#1,\mathrm{str}}}
\newcommand{\magdescriptor}[1]{\descriptor^{#1,\mathrm{mag}}}
\newcommand{\Othree}{\mathrm{O}(3)}

\begin{document}

\title{NEP-Spin: A spin-dependent neuroevolution potential for large-scale spin--lattice simulations}

\author{Chengbing Chen}
\affiliation{Graduate School of China Academy of Engineering Physics, Beijing 100193, P. R. China}
\author{Zheyong Fan}
\affiliation{College of Physical Science and Technology, Bohai University, Jinzhou 121013, P. R. China}
\author{Ben Xu}
\email{bxu@gscaep.ac.cn}
\affiliation{Graduate School of China Academy of Engineering Physics, Beijing 100193, P. R. China}

\date{\today}

\begin{abstract}
NEP-Spin is a \glsentrylong{nep} for coupled atomic motion and variable-length
magnetic moments. Local bond and spin tensors give an energy invariant under
independent spin and spatial rotations without \glsentrylong{soc} and under
joint rotations with it. We benchmark energies, atomic forces, and magnetic
forces on public datasets for collinear Fe--Al and noncollinear Fe.
In multiferroic \glsentrylong{bfo}, the predicted
zero-field cycloid period is 69.7~nm, compared with 62--63~nm from neutron
scattering. An electric field of 20~mV/\AA{} opposite to the polarization lengthens the
period to 93.2~nm, while imposing the same polar displacement at zero field
gives 85.1~nm.
In a uniform antiferromagnetic cell at fixed lattice vectors, 6~T magnetic fields
along orthogonal basal axes reorient the Fe moments and give opposite signs
of the ionic polarization change $\Delta P_y$. First-principles ionic
relaxations at the predicted moment vectors reproduce both signs.
In a 2.79-million-atom ferroelectric twin, the cycloid rotation planes
tilt by different amounts from their shared initial plane and remain distinct
over 10~ps at 100~K. At 10~ps the propagation axes lie within about
$7^\circ$ of their initial directions.
With inertial spin dynamics, GPUMD advances a 5.93-million-atom
BiFeO$_3$ cell at 21.8 million atom-steps per second on four RTX 4090 GPUs.
\end{abstract}

\maketitle

\section{Introduction}
\label{sec:introduction}

Coupling between atomic motion and local magnetic moments influences the
magnetic structure and finite-temperature behavior of magnetic materials
\cite{Ma2008Largescale}.  Atomistic descriptions of this coupling treat both
atomic positions $\Rset=\{\rvec_i\}$ and local magnetic moments
$\Sset=\{\spinvec_i\}$ as degrees of freedom.  Magnetic machine-learning potentials approximate the resulting energy surface
from first-principles data
\cite{Novikov2022Magnetic,Yu2024Physicsinformed,Yang2024Deep}.
At fixed atomic positions and in zero field, isotropic exchange is
invariant under a global rotation of all moments. Rotating a single
moment can change the exchange energy by changing its alignment with
its neighbors. \Gls{soc} also makes the energy
sensitive to the orientation of the moments relative to the lattice.

Spin-dependent atom-centered symmetry
functions, \glspl{mmtp}, and vector-field spectra encode
magnetic moments and their coupling to the local atomic environment
\cite{Eckhoff2021Highdimensional,Novikov2022Magnetic,Domina2022Spectral}.
Magnetic atomic cluster expansion extends this construction through a
systematically truncated basis, including a noncollinear model for iron
\cite{Drautz2020Atomic,Rinaldi2024Noncollinear}.  Graph-based and equivariant
models provide further routes to representing coupled spin--lattice energy
surfaces, with extensions that account for \gls{soc}
\cite{Yu2024Spindependent,Yu2024Physicsinformed,Yang2024Deep,
Ho2026Equivariant,Gao2026STEP,Yuan2024MagNet,Wang2026HotPPSpin}.
Applications have addressed magnetic states, magnon spectra, and coupled
spin--lattice behavior
\cite{Chapman2022Machinelearned,Yu2024Physicsinformed,Yuan2024MagNet,Gao2026STEP}.
Alongside energies, atomic forces, and stresses, magnetic forces have also
been used as training targets, improving the reliability of magnetic
relaxation in Fe--Al \cite{Kotykhov2024Fitting}.

Beyond aggregate fitting errors, comparisons of exchange,
\gls{dmi}, magnetic anisotropy, and
magnetovolume coupling assess the energy changes relevant to magnetic
applications.  Effective exchange parameters extracted from local
rotations depend on the reference state and spin-model
convention~\cite{Streib2022Adiabatic}.

Here we extend the \gls{nep} to variable-length local
moments, retaining its short-range descriptors, atomic neural networks,
and implementation on \glspl{gpu}~\cite{Fan2021Neuroevolution}. Drawing on angular-momentum
representations of local vector fields~\cite{Domina2022Spectral}, we expand
bond directions and moments separately. The symmetry of the energy determines how
we contract these tensors. Without \gls{soc}, the spatial and spin indices transform
independently. With \gls{soc}, the two transform in a common frame, allowing
the atomic energy to depend on the orientation of the moments relative
to nearby bonds.

Two public datasets probe different derivatives of the spin-dependent
energy. Collinear Fe--Al provides magnetic forces along a single spin axis,
whereas noncollinear Fe provides all three Cartesian components. Rhombohedral \gls{bfo} provides a test of \gls{soc} because
exchange, \gls{dmi}, and anisotropy compete and depend on the surrounding
oxygen coordinates~\cite{Xu2019Magnetic}. Its magnetic response to a
structural distortion probes the joint dependence of the energy on
positions and moments.

\section{Methods}
\label{sec:tensor-framework}

\subsection{Magnetic descriptors}

NEP-Spin retains the structural \gls{nep} descriptors and augments the input
to each atomic network with scalars constructed from the local moments.
At each site we sum contributions from the central moment and its
neighbors into tensors, grouping terms with the same transformation properties. The
squared norm of each sum contains cross terms between sites. For
example, a vector sum of neighboring moments retains their relative
alignment through the dot products in its squared norm. Bond directions
and radial weights make these correlations sensitive to the atomic
geometry. We use $C$ radial channels, retain angular ranks up to $L$,
and expand the moments in polynomials of degree up to $S$.

Let $\rvec_{ik}$ point from atom $i$ to a periodic image of atom $k$,
$r_{ik}=|\rvec_{ik}|$, and $\hat\rvec_{ik}=\rvec_{ik}/r_{ik}$.
Every atom is an energy center. Only the selected magnetic species carry
independent Cartesian moments $\spinvec_i$, measured in $\mu_{\rm B}$. For other
species the effective moment is zero. Such an atom can still contribute
magnetic energy through the moments of its neighbors, and its displacement
can change the magnetic environment of another center.

Each bond inside the magnetic cutoff carries $C$ radial weights. For channel
$c$ the weight is
\begin{equation}
 w_{ikc}=\sum_n A_{nc}\sum_b h^{Z_i Z_k}_{nb}\,
             f_b(r_{ik};r^{\rm s}_{ik}).
 \label{eq:compressed-radial}
\end{equation}
Here $Z_i$ is the species of atom $i$, $h^{Z_i Z_k}_{nb}$ are the radial
coefficients of the angular branch of structural \gls{nep}, and $A_{nc}$ projects
these functions onto $C$ learned channels. We fit $h$ and $A$ jointly.
The \gls{bfo} training also includes a network-optimization stage in which
$h$ and $A$ are held fixed.
The magnetic branch evaluates the basis at the pair cutoff
$r^{\rm s}_{ik}=(r^{\rm s}_{Z_i}+r^{\rm s}_{Z_k})/2$ (Appendix~\ref{app:grammar}).
All tensor couplings in channel $c$ use the same $w_{ikc}$.

The bond direction is expanded in real spherical harmonics
$Y_{gm}(\hat\rvec_{ik})$ of rank $0\le g\le L$. To retain the moment magnitude
as well as its direction, we use the real solid harmonic
$\mathcal Y_{ja}(\spinvec)=|\spinvec|^jY_{ja}(\hat{\spinvec})$ and form
homogeneous polynomials
\begin{equation}
 \Phi_{dja}(\spinvec)=(\spinvec\cdot\spinvec)^{(d-j)/2}
                        \mathcal Y_{ja}(\spinvec).
 \label{eq:spin-primitives}
\end{equation}
Here $j$ is the spin rank and $a$ labels its $2j+1$ real components. The degree
$d$ runs from zero to $S$. At each degree we retain the ranks
$0\le j\le\min(d,L)$ for which $d-j$ is even. For example, degree one
gives the moment vector up to normalization, while degree two gives its
squared magnitude and, for $L\ge2$, its traceless quadrupole. In Cartesian
form these polynomials and their derivatives are defined even at zero
moment. The constant $\Phi_{000}=1$ supplies a purely geometric contribution.

Without \gls{soc}, a rigid rotation of the spins at fixed lattice must leave
the energy unchanged. We therefore retain both component indices of the
product $Y_{gm}\Phi_{dja}$ and sum contributions at fixed $(g,j)$,
\begin{align}
 T^{ic}_{gj,ma}
 &=T^{ic,\mathrm{on}}_{gj,ma}
 +\sum_{k\in\mathcal N_i} w_{ikc}Y_{gm}(\hat\rvec_{ik})\nonumber\\
 &\qquad{}\times\sum_{d,t} B^{gj}_{dtc}\,\Phi_{dja}(\spinvec_{t(i,k)}).
 \label{eq:product-density}
\end{align}
Here $\mathcal N_i$ contains the neighbor images within the magnetic cutoff.
The inner sum includes every degree admitting rank $j$ and either
$t(i,k)=i$ or $k$, with independent learned weights $B^{gj}_{dtc}$.
For $d=0$ the constant is counted once per bond. The on-site term contributes only at $g=0$. Independent spatial
and spin rotations act on the indices $m$ and $a$, respectively. The normalized squared Frobenius norm
$\sum_{m,a}(T^{ic}_{gj,ma})^2/[(2g+1)(2j+1)]$ is invariant under both
rotations. The network also receives the signed scalar $T^{ic}_{00,00}$
and, once per atom across all channels, each local scalar $\Phi_{d00}(\spinvec_i)$ of positive even degree.

Exchange survives this independent rotation symmetry. For example, select
the degree-one contribution at $(g,j)=(0,1)$. Its on-site and
neighbor terms have the Cartesian form
$u_c\spinvec_i+v_c\sum_k w_{ikc}\spinvec_k$, with harmonic normalization absorbed
into $u_c$ and $v_c$. Squaring this vector produces the cross term
$2u_cv_c\sum_k w_{ikc}\spinvec_i\cdot\spinvec_k$. A displacement changes the
weights, and rotating a single moment changes the dot products.
The squared neighbor sum also contains terms proportional to
$w_{ikc}w_{ik'c}\spinvec_k\cdot\spinvec_{k'}$ for distinct $k,k'$.
These terms survive on a center with no independent moment. Moving such
a center can therefore change the interaction between two magnetic
neighbors even when the lattice cannot select a preferred spin axis.

With \gls{soc}, a joint rotation of the lattice and moments preserves the
energy, while rotating the moments alone can change it. We retain the
radial weights and spin polynomials of the isotropic construction.
In the tensor sum, spatial and spin indices now couple
to a total angular momentum. Contributions with the same total rank and
parities can interfere in the squared norm even when their separate
spatial and spin ranks differ.

A displacement is a polar vector and a magnetic moment is an axial vector.
Under a common orthogonal transformation $\bm Q$ they obey
\begin{equation}
 \rvec_{ik}\mapsto \bm Q\rvec_{ik},\qquad
 \spinvec_i\mapsto\det(\bm Q)\bm Q\spinvec_i .
 \label{eq:joint-o3}
\end{equation}
In particular, inversion multiplies a bond harmonic of rank $g$ by
$(-1)^g$ and leaves the moment unchanged. Time reversal flips every
moment and multiplies a degree-$d$ primitive by $(-1)^d$. These two
signs determine which contributions can mix before forming a scalar.

The contribution from
neighbor $k$ couples its bond harmonic to a spin primitive on either end
of the bond. In real Clebsch--Gordan notation,
\begin{equation}
 [Y_g\otimes\Phi_{dj}]_{JM}
 =\sum_{m,a}\langle gm,ja\mid JM\rangle\,Y_{gm}\Phi_{dja},
 \label{eq:cg}
\end{equation}
where $|g-j|\le J\le\min(g+j,L)$.
In a path $\alpha=(g,d,j,t,J)$, the bond harmonic of rank $g$ couples
to a spin polynomial of degree $d$ and rank $j$ to give total rank $J$.
The site index $t$ selects the central or neighboring moment.
For $d=0$ the constant is included only once per bond.
Paths with the same total rank $J$, inversion parity $p=(-1)^g$,
and time-reversal parity $\eta=(-1)^d$ accumulate in one density,
\begin{align}
 H^{ic}_{Jp\eta,M}
 &= H^{ic,\mathrm{on}}_{Jp\eta,M}+
 \sum_{k\in\mathcal N_i}w_{ikc}\!
 \sum_{\alpha\to(J,p,\eta)} B_{\alpha c}\nonumber\\
 &\qquad{}\times [Y_g(\hat\rvec_{ik})\otimes
 \Phi_{dj}(\spinvec_{t(i,k)})]_{JM}.
 \label{eq:soc-density}
\end{align}
$B_{\alpha c}$ is a learned coefficient shared across bonds. The on-site
term contributes to sectors $(j,+1,(-1)^d)$ through independently weighted
positive-degree primitives of the central moment.

Each sector supplies a scalar,
$\sum_M(H^{ic}_{Jp\eta,M})^2/(2J+1)$. The network also receives the signed value of the inversion- and time-even
scalar $H^{ic}_{0++,0}$ and the on-site scalars
$\Phi_{d00}(\spinvec_i)$ for positive even $d$, each on-site scalar counted
once. Because paths are summed before squaring, their cross terms survive. For
example, the $g=j=2$, $J=0$ path on the central moment is proportional to
\begin{equation}
 \sum_k w_{ikc}
 \left[(\hat\rvec_{ik}\cdot\spinvec_i)^2-\tfrac13|\spinvec_i|^2\right].
 \label{eq:anisotropy-channel}
\end{equation}
This scalar can change when the moment turns relative to an anisotropic bond
environment. In a shell with an isotropic weighted second moment, the
sum vanishes.

A different contraction illustrates antisymmetric exchange. For
$g=j=d=1$ and $J=0$, a bond and a moment couple to their dot product.
The center and neighbor paths in the $(0,-,-)$ sector therefore contain,
up to normalization, $\vect P_c\cdot\spinvec_i$ and
$\sum_k w_{ikc}\hat\rvec_{ik}\cdot\spinvec_k$, respectively, where
$\vect P_c=\sum_k w_{ikc}\hat\rvec_{ik}$.
Their cross term in the sector norm contains
$(\vect P_c\cdot\spinvec_i)(\hat\rvec_{ik}\cdot\spinvec_k)$.
The part antisymmetric in the two spin indices is proportional to
\begin{equation}
 (\vect P_c\times\hat\rvec_{ik})\cdot(\spinvec_i\times\spinvec_k).
 \label{eq:chiral-channel}
\end{equation}
The geometry supplies the axial vector required for a \gls{dmi}-like bilinear
interaction. This contribution vanishes if $\vect P_c$ is zero or parallel
to the bond. The effective \gls{dmi} for an ordered pair depends on the sum of
site energies and on the symmetry of the full environment.

\subsection{Energy derivatives and training}

For each atom we concatenate the structural \gls{nep} descriptors
$\structdescriptor{i}$ with the scalar density contractions and local
spin scalars $\magdescriptor{i}$. For the combined vector
$\descriptor^i=[\structdescriptor{i},\magdescriptor{i}]$, the \gls{nep}
readout~\cite{Fan2021Neuroevolution} is
\begin{align}
 U &=\sum_i\left(E^0_{Z_i}+U_i\right)+U_{\rm ZBL}+U_{\rm wall},\nonumber\\
 U_i &= \sum_\mu w^{(1,Z_i)}_\mu
 \tanh\!\left(\sum_\nu w^{(0,Z_i)}_{\mu\nu}a_\nu q^i_\nu
                      -b^{(0,Z_i)}_\mu\right)-b^{(1)} .
 \label{eq:scalar-energy}
\end{align}
$E^0_{Z_i}$ is a species reference energy, $a_\nu$ is a fixed descriptor
scale, and $b^{(1)}$ is a shared bias. The optional Ziegler--Biersack--Littmark repulsion
$U_{\rm ZBL}$ is spin independent~\cite{Xu2025GPUMD}. Relative bond vectors ensure translation invariance, and summing over
neighbors preserves invariance under permutations of equivalent atoms. The scalar inputs
are even under global moment reversal, so the zero-field energy has this
symmetry for any network weights. At vanishing moments the degree-zero
bond densities persist, and the model reduces to a geometry-dependent
energy. On an isolated atom only the on-site primitives contribute to the
tensor densities.

Outside the sampled range of moment magnitudes, the predicted longitudinal
restoring force can weaken or reverse sign. The total energy can therefore
include a species-dependent longitudinal boundary potential,
\begin{align}
 U_{\rm wall} &= \sum_i \Bigg[
 \frac{k^{\rm hi}_{Z_i}}{3}
 \max\!\left(0,\, |\spinvec_i| - s^{\rm hi}_{Z_i}\right)^3 \nonumber\\
 &\qquad
 + \frac{k^{\rm lo}_{Z_i}\,s^{\rm lo}_{Z_i}}{3}
 \max\!\left(0,\, 1 - \frac{|\spinvec_i|^2}{(s^{\rm lo}_{Z_i})^2}\right)^3
 \Bigg] .
 \label{eq:moment-wall}
\end{align}
For each species $Z_i$, the upper and lower thresholds are chosen above and
below its sampled range of moment magnitudes, respectively. Each active
threshold is positive, with $s^{\rm lo}_{Z_i}<s^{\rm hi}_{Z_i}$ when both
walls are active. The nonnegative coefficients $k^{\rm hi}_{Z_i}$ and
$k^{\rm lo}_{Z_i}$ set the respective stiffnesses. A term with zero
stiffness is omitted. Both terms are disabled by default and vanish
identically for species without active spin degrees of freedom.
Each term is $C^2$ at its threshold. Within the sampled range, the wall
energy and its first and second derivatives vanish, leaving equilibrium
moments in that range and their lattice dependence governed by the network.

Above the upper threshold, the wall exerts the inward magnetic force
$-k^{\rm hi}_{Z_i}(|\spinvec_i|-s^{\rm hi}_{Z_i})^2\hat{\spinvec}_i$.
Below the lower threshold, the force is
$2k^{\rm lo}_{Z_i}[1-|\spinvec_i|^2/(s^{\rm lo}_{Z_i})^2]^2
\spinvec_i/s^{\rm lo}_{Z_i}$. Both forces follow from the same scalar
energy and are purely longitudinal, so
$\spinvec_i\times\mforce_i^{\rm wall}=\bm 0$.
Equation~\eqref{eq:moment-wall} depends only on moment magnitudes. It
therefore preserves $\Othree$ and time-reversal symmetry and contributes
neither atomic forces nor virials. The lower term is smooth in
$|\spinvec_i|^2$, so its magnetic force vanishes at $\spinvec_i=\bm 0$.
The lower wall is intended only for exploratory sampling in which small
moments represent unphysical collapse. It is kept disabled when studying
longitudinal moment suppression, demagnetization, or magnetic phase transitions.
Both walls are disabled in the Fe, Fe--Al, and \gls{bfo} models used in this study.
Figure~\ref{fig:bfo-eos}(c) illustrates the effect of adding an upper
wall to the \gls{bfo} model in a fixed-geometry scan.

The atomic force, magnetic force on an active site, and spin torque are
\begin{equation}
 \force_k=-\left.\frac{\partial U}{\partial\rvec_k}\right|_{\Sset},
 \qquad
 \mforce_i=-\left.\frac{\partial U}{\partial\spinvec_i}\right|_{\Rset},
 \qquad
 \torque_i=\spinvec_i\times\mforce_i .
 \label{eq:forces}
\end{equation}
The magnetic force, in eV/$\mu_{\rm B}$, is conjugate to a change of the full
moment vector. Excluded spin species have zero magnetic force by
convention. All atoms receive the positional derivative of the full
energy.
Along a prescribed spin rotation at fixed geometry, the generalized torque is
$-\mathrm dU/\mathrm d\theta=\sum_i\mforce_i\cdot
\mathrm d\spinvec_i/\mathrm d\theta$.
Where $U$ is twice differentiable, the same mixed derivative
governs the response in both directions,
\begin{equation}
 \frac{\partial F_{k,\alpha}}{\partial s_{i,\beta}}=
 -\frac{\partial^2 U}{\partial s_{i,\beta}\,\partial r_{k,\alpha}}=
 \frac{\partial M_{i,\beta}}{\partial r_{k,\alpha}} .
 \label{eq:mixed-response}
\end{equation}
For an infinitesimal homogeneous deformation
$\rvec'=(\bm I+\delta\bm\xi)\rvec$, the dimensionless displacement
gradient $\delta\bm\xi$ defines the configurational virial $\bm W$ through
$\delta U=-\sum_{a,b}W_{ab}\,\delta\xi_{ab}$ at fixed Cartesian magnetic
moments. The symmetric part of $\delta\bm\xi$ strains the lattice, while
the antisymmetric part rotates it relative to the moments. With \gls{soc},
this relative rotation can change the energy of anisotropic spin--lattice
interactions, including magnetocrystalline anisotropy and anisotropic
exchange. The full $3\times3$ virial can therefore have unequal
off-diagonal entries. In zero field, invariance under simultaneous rotation
of the lattice and moments relates their differences to the net magnetic
torque,
\begin{equation*}
 W_{ab}-W_{ba}=\sum_i(s_{i,a}M_{i,b}-s_{i,b}M_{i,a}) .
\end{equation*}

For a symmetric strain
$\delta\bm\varepsilon=(\delta\bm\xi+\delta\bm\xi^{\mathsf T})/2$,
the energy variation is $\delta U=-\sum_{a,b}W^{\rm sym}_{ab}\,
\delta\varepsilon_{ab}$, with
\begin{equation}
 \bm W^{\rm sym}=\tfrac12(\bm W+\bm W^{\mathsf T}),\qquad
 \bm\sigma^{\rm sym}=-\bm W^{\rm sym}/V ,
 \label{eq:sym-virial}
\end{equation}
where $\bm\sigma^{\rm sym}$ is the symmetric, tensile-positive
configurational stress. For engineering shear $\gamma_{xy}=2\varepsilon_{xy}$,
the corresponding derivative is
$-\partial U/\partial\gamma_{xy}=(W_{xy}+W_{yx})/2$.

We optimize the density coefficients and atomic-network parameters jointly
with the \gls{snes}~\cite{Schaul2011SNES}. Magnetic-force
labels constrain the energy cost of varying a moment in both magnitude
and direction. Full virial labels constrain the response to deformation
at fixed moments, including relative rotations. Symmetric virial or stress
labels constrain the symmetric strain response. The data loss weights the
\gls{rmse} for each labeled quantity,
\begin{equation}
 \mathcal L_{\rm data}=\lambda_E R_E+\lambda_F R_F+\lambda_V R_V
                 +\mathcal L_M+\lambda_\tau R_\tau .
 \label{eq:data-loss}
\end{equation}
The optimizer also includes \gls{nep} parameter regularization~\cite{Fan2021Neuroevolution}.
$R_E$ compares energies per atom across structures after subtracting the mean per-atom
energy residual in the training batch, weighted by the energy-label weights. $R_F$ averages the squared
Cartesian force residuals over three components per atom before taking
the square root. The formulas below give the unit-configuration-weight
case. During fitting, a configuration weight multiplies its residuals
before squaring. We report unweighted \gls{rmse} values for all quantities.

For a labeled active moment let $\Delta\mforce_i=\mforce_i-\mforce_i^{\rm ref}$.
The Cartesian magnetic-force objective and the optional torque \gls{rmse} are
\begin{align}
 \mathcal L_M &= \lambda_M R_M,\qquad
 R_M^2=\frac{1}{3N_M}\sum_{i\in\mathcal A}|\Delta\mforce_i|^2,\nonumber\\
 R_\tau^2&=\frac{1}{2N_\tau}\sum_{i\in\mathcal A_\tau}
                    |\spinvec_i\times\Delta\mforce_i|^2 .
 \label{eq:mag-loss}
\end{align}
$\mathcal A$ contains the $N_M$ active moments on configurations with
magnetic-force labels, including moments of zero magnitude.
The optional torque term penalizes the part of the magnetic-force error that produces a torque.
Appendix~\ref{app:loss-options} gives its site selection and the alternative
loss for separately weighted magnetic-force components.

For a labeled structure $n$ of $N_n$ atoms and volume $V_n$, let
$\bm w_n=\bm W_n/N_n$ and $\bm w_n^{\rm sym}=\bm W_n^{\rm sym}/N_n$ denote
the predicted full and symmetric per-atom virials, respectively. The
reference $\bm w_n^{\rm ref}$ uses the same index convention. A
tensile-positive stress reference is converted to a per-atom virial by
multiplication by $-V_n/N_n$. For a full virial reference, we compare all
nine Cartesian entries, $\mathcal I_n=\{x,y,z\}^2$, with residuals
$\Delta w_{n,ab}=w_{n,ab}-w_{n,ab}^{\rm ref}$. For a symmetric virial or
stress reference, we compare the six independent entries,
$\mathcal I_n=\{(x,x),(y,y),(z,z),(x,y),(y,z),(z,x)\}$, with residuals
$\Delta w_{n,ab}=w_{n,ab}^{\rm sym}-w_{n,ab}^{\rm ref}$.

The virial loss uses
\begin{equation}
 R_V^2=\frac{1}{N_{\rm comp}}\sum_{n\in\mathcal V}
 \sum_{(a,b)\in\mathcal I_n}c_{ab}^2(\Delta w_{n,ab})^2 .
 \label{eq:virial-loss}
\end{equation}
Here $\mathcal V$ contains the labeled structures and
$N_{\rm comp}=\sum_{n\in\mathcal V}|\mathcal I_n|$ counts the supervised
components. During training, $c_{aa}=1$ and $c_{ab}=\lambda_{\rm shear}$
for $a\ne b$. All component weights are set to one when reporting the
virial \gls{rmse}.

\subsection{Datasets and computational details}
\label{sec:computational-settings}

The Fe--Al dataset of Kotykhov \textit{et al.}~\cite{Kotykhov2024Fitting}
contains collinear moments and longitudinal magnetic forces on both Fe and Al.
The 2024 release retains the 2,012 training configurations from the 2023
study~\cite{Kotykhov2023Constrained} and adds 620 configurations. It supplies
magnetic-force labels for all 2,632 configurations. Our split used 1,981 configurations for fitting, 365 for validation,
and 286 for testing. For comparison with Table~II of
Ref.~\cite{Kotykhov2024Fitting}, we evaluated the final model on all 2,632.
For elemental Fe we used the noncollinear reference
calculations of Zheng \textit{et al.}~\cite{Zheng2026Fe}, obtained with a
\gls{dzp} basis of numerical atomic orbitals.
We fitted all 34,327 Fe configurations and report the training errors
for comparison with the published DeePSPIN-DZP training errors.

Both fits used $C=3$, $L=S=2$, 30 structural descriptors, and a hidden
layer of 30 neurons. For Fe--Al we used the 31 magnetic descriptors
invariant under independent spin and spatial rotations. For Fe we used
the 34 magnetic descriptors invariant under joint rotations, allowing
orientation dependence relative to the lattice. Each fit ran for
100,000 \gls{snes} generations with weights of 1, 1, 0.1, and 1 for energy,
atomic force, virial, and Cartesian magnetic force, respectively.

The \gls{bfo} reference structures were constructed from the rhombohedral
$R3c$ phase.  Non-diagonal expansions of the ten-atom primitive cell
provided 40- and 50-atom supercells, containing four and five primitive
cells, respectively.  NepTrain~\cite{Chen2025NepTrain} was used to automate
successive cycles of potential fitting, \gls{md} sampling,
representative-configuration selection, and labeling with \gls{dft}. The magnetic
degrees of freedom are the Fe moments. The response configurations sampled
Fe rotations and structural perturbations.
The \gls{soc} model used three radial channels ($C=3$), a maximum angular rank $L=2$,
spin-polynomial degree $S=2$, and a single hidden layer of 30 neurons.
The radial and angular structural descriptors have cutoffs of 6 and
5~\AA{}, respectively. The magnetic descriptors have a cutoff of 6~\AA{}.

\gls{dft} reference calculations were performed using
VASP~\cite{Kresse1994AbInitio,Kresse1996Efficient} with the
projector-augmented-wave method~\cite{Blochl1994Projector,Kresse1999PAW}
and the \gls{pbe} exchange--correlation
functional~\cite{Perdew1996Generalized}.  The Bi $5d^{10}6s^26p^3$,
Fe $3d^74s^1$, and O $2s^22p^4$ electrons were treated explicitly as
valence electrons.  The localized Fe $3d$ states were described using the
Dudarev \gls{dft}+$U$ approach~\cite{Dudarev1998LSDAU}, with
$U_{\mathrm{eff}}=U-J=4$~eV.  For the 1,133 reference configurations in the BFO dataset, the plane-wave cutoff was 800~eV, the Brillouin zone was sampled
using $\Gamma$-centered meshes with a reciprocal-space spacing parameter
of $0.25$~\AA$^{-1}$, and the electronic self-consistency threshold was
$10^{-7}$~eV.  Noncollinear magnetism and \gls{soc} were included.
The DeltaSpin method~\cite{Cai2023DeltaSpin} constrained the Fe local
magnetic moments in both direction and magnitude, while the Bi and O
magnetization was allowed to respond self-consistently.  The resulting
labels comprised energies, atomic forces, virials, local magnetic moments,
and the magnetic forces conjugate to the constrained moments.
The magnetic-force sign convention follows Eq.~\eqref{eq:forces}.
The \gls{dft} reference for the volume and moment scans was a ten-atom cell
relaxed on the basal G-type \gls{afm} branch. These scans
used an electronic threshold of $10^{-10}$~eV. The relaxed reference
volume was 128.238~\AA$^3$.

Starting from the relaxed ten-atom reference cell, we varied the volume
at fixed Fe moments and the moment magnitudes at fixed geometry.
We also varied both quantities in a crossed scan to measure their coupling.
We applied the same constraints and energy normalization to the
NEP-Spin and \gls{dft} scans.

For the ordered nearest-neighbor Fe pair in Fig.~\ref{fig:bfo-magnetic}(a)
(4.00~\AA{} apart in the reference geometry), we extracted effective exchange coefficients
from four total energies with independently signed $5^\circ$ transverse
rotations about a G-type \gls{afm} reference. For transverse components
$\alpha,\beta\in\{y,z\}$ on sites $i,j$, let $E_{\zeta_i\zeta_j}$ denote
the energy with transverse projections $\zeta_i\sin\theta$ and $\zeta_j\sin\theta$
of the two unit moments, where $\zeta_i,\zeta_j=\pm1$ and $\theta=5^\circ$.
From $B_{\alpha\beta}=\sum_{\zeta_i,\zeta_j}\zeta_i\zeta_j E_{\zeta_i\zeta_j}/(4\sin^2\theta)$
we obtain
$J_\perp=(B_{yy}+B_{zz})/2$ and
$D_x=(B_{yz}-B_{zy})/2$ for the same ordered pair in \gls{dft} and NEP-Spin.
We measured the magnetocrystalline anisotropy by rotating the
collinear \gls{afm} axis at fixed ionic positions and moment magnitudes.

For the polarization comparison in Fig.~\ref{fig:bfo-smallcell-field},
we selected one zero-field state and nine finite-field states from the
NEP-Spin Zeeman scan. In each \gls{dft} calculation, DeltaSpin held both Fe moment
vectors at their NEP-Spin values while all ions relaxed at fixed lattice
vectors. We used a plane-wave cutoff of 800 eV and a reciprocal-space k-point spacing parameter of 0.25 \AA$^{-1}$, with an
electronic energy tolerance of $10^{-10}$~eV, a moment tolerance of
$10^{-8}\,\mu_{\rm B}$, and a force tolerance of $10^{-5}$~eV/\AA.
For each method, the ionic polarization change was calculated as
$\Delta P_y=(e/V)\,\hat{\vect Y}\cdot\sum_i\vect Z_i^*\Delta\vect r_i$,
where $\Delta\vect r_i$ is the displacement from that method's relaxed
zero-field geometry, $V$ is the cell volume, and the same Born
effective charge tensors $\vect Z_i^*$ were used for both methods.

\section{Results and discussion}
\label{sec:results}

\subsection{Public benchmarks}
\begin{table*}[htbp]
\caption{\gls{rmse} on the public Fe--Al and Fe datasets. Both Fe--Al rows cover
the same 2,632 configurations. The \gls{mmtp} values are means over five fits with the reported $2\sigma$ intervals ($N_\psi=5$, $w_t=0.1\,\mu_{\rm B}^2$).
For Fe, DeePSPIN-DZP reports training errors, and STEP-full reports errors
over the original dataset after a 90:10 training--validation split
(Table~III of Ref.~\cite{Gao2026STEP}).
Errors in the \glsentrylong{mf} (\glsentryshort{mf}) use the common spin-axis component for Fe--Al
and all three Cartesian components for Fe. Dashes indicate unreported values.}
\label{tab:public-accuracy}
\small
\begin{tabular*}{\linewidth}{@{\extracolsep{\fill}}llrrrr@{}}
\toprule
System & Model & $U$ (meV/atom) & $F$ (meV/\AA) & Stress (GPa) & \glsentryshort{mf} (meV/$\mu_{\rm B}$)\\
\midrule
Fe--Al & \gls{mmtp}~\cite{Kotykhov2024Fitting} & $1.73\pm0.06$ & $64.40\pm3.95$ & $0.469\pm0.029$ & $15.2\pm0.4$\\
       & NEP-Spin & 1.41 & 24.8 & 0.220 & 24.43\\
\midrule
Fe & DeePSPIN-DZP~\cite{Zheng2026Fe} & 6.88 & 106.0 & --- & 28.6\\
   & STEP-full~\cite{Gao2026STEP} & 1.283 & 46.368 & --- & 6.974\\
   & NEP-Spin & 5.34 & 109.5 & 0.697 & 16.72\\
\bottomrule
\end{tabular*}
\end{table*}

\begin{figure*}[htbp]
\centering
\includegraphics[width=\linewidth]{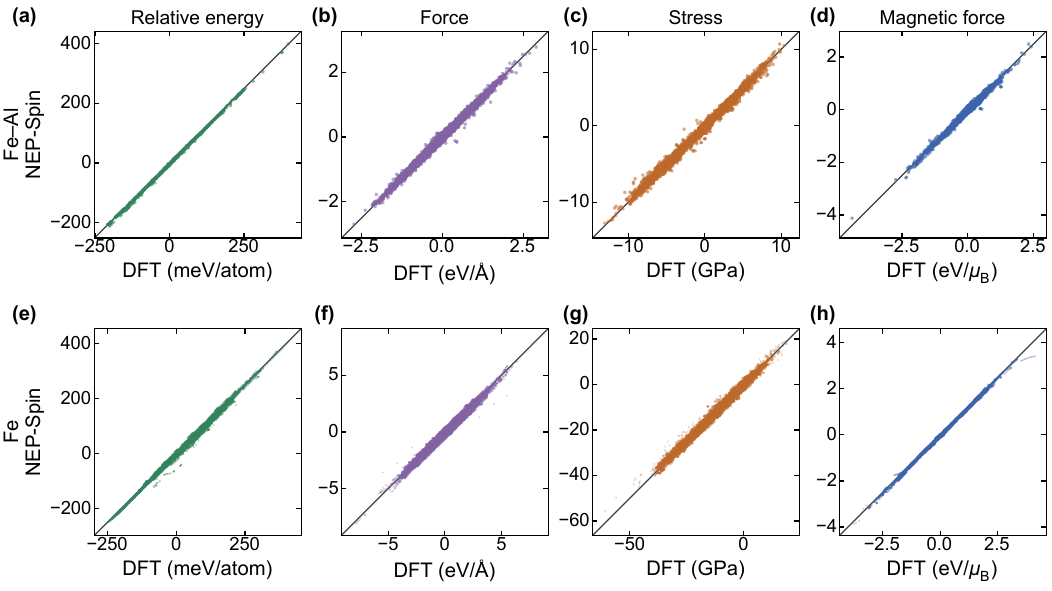}
\caption{NEP-Spin predictions against \gls{dft} for all 2,632 Fe--Al configurations
(top) and all 34,327 Fe configurations (bottom).
Columns show energy per atom relative to the same elemental references
on both axes, Cartesian atomic forces, tensile-positive stress, and magnetic
forces. Each Fe--Al atom contributes one
longitudinal magnetic-force component; each Fe atom contributes three
Cartesian components. The diagonal indicates equality.}
\label{fig:public-test-parity}
\end{figure*}

On the 2024 Fe--Al dataset, NEP-Spin gives smaller energy, atomic-force,
and stress errors than the \gls{mmtp} ensemble (Table~\ref{tab:public-accuracy}).
The longitudinal magnetic-force error is larger for NEP-Spin,
24.43 versus $15.2\pm0.4$~meV/$\mu_{\rm B}$
(Fig.~\ref{fig:public-test-parity}).
On the 286 held-out Fe--Al configurations, NEP-Spin gives errors of
1.41~meV/atom in energy, 24.3~meV/\AA{} in atomic force, 0.227~GPa in
stress, and 26.76~meV/$\mu_{\rm B}$ in longitudinal magnetic force.
For noncollinear Fe, the NEP-Spin Cartesian magnetic-force \gls{rmse} is 16.72~meV/$\mu_{\rm B}$, compared with
28.6~meV/$\mu_{\rm B}$ for DeePSPIN-DZP (training) and 6.974~meV/$\mu_{\rm B}$ for STEP-full. For NEP-Spin the training error per transverse
component is 13.1~meV/$\mu_{\rm B}$.
The atomic-force error of NEP-Spin (109.5~meV/\AA) is close to that of
DeePSPIN-DZP (106.0~meV/\AA).

\subsection{Spin--lattice coupling in \texorpdfstring{BiFeO$_3$}{BiFeO3}}
\label{sec:bfo-potential}

\begin{figure*}[htbp]
\centering
\includegraphics[width=\textwidth]{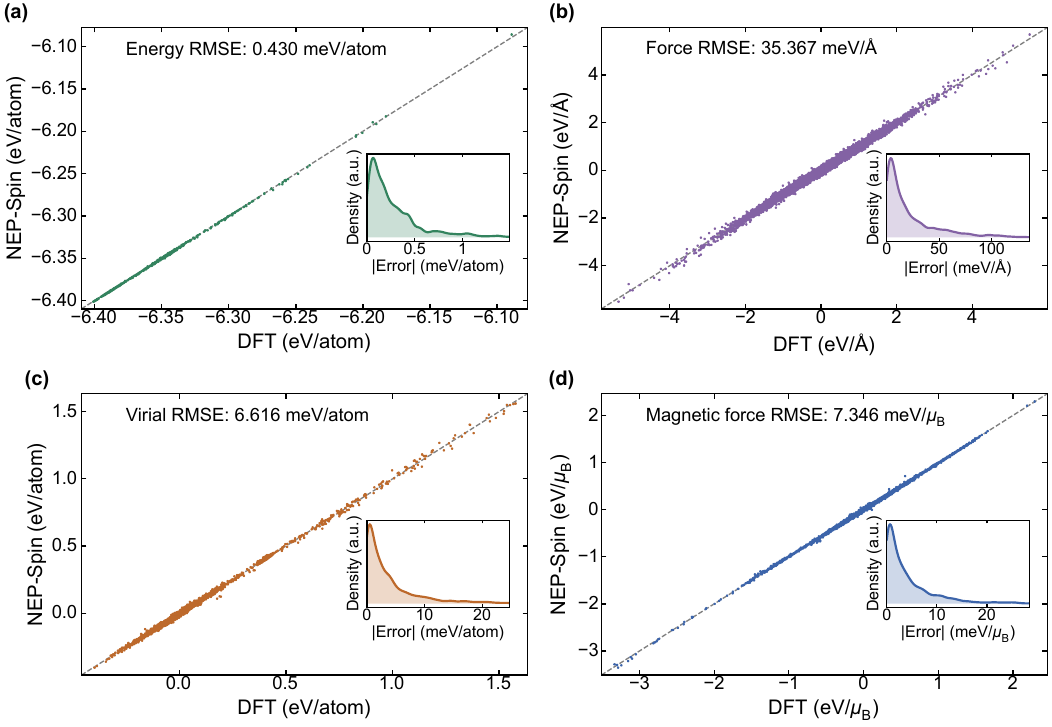}
\caption{NEP-Spin predictions against \gls{dft} for the 1,355 base training configurations
of \gls{bfo}. Panels show energy per atom, Cartesian atomic forces,
virials per atom, and Cartesian magnetic forces on Fe. Insets show the
absolute-error distributions.}
\label{fig:bfo-parity}
\end{figure*}

\begin{figure*}[htbp]
\centering
\includegraphics[width=\linewidth]{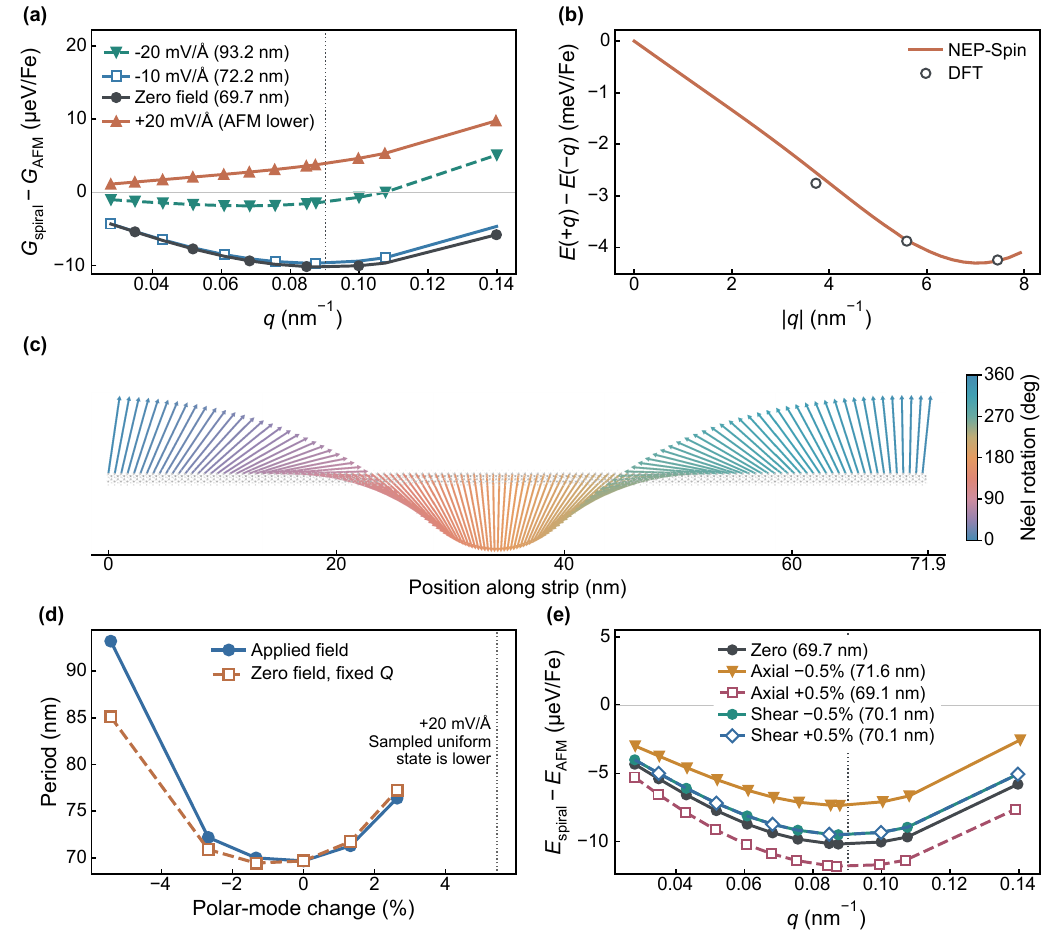}
\caption{Cycloid energetics and structure in \gls{bfo}. (a) Electric enthalpy
of relaxed one-turn spirals relative to the uniform \gls{afm} state at the same
field. The dotted line marks the fitted zero-field minimum.
(b) Fixed-geometry handedness splitting at finite $|q|$ from NEP-Spin
and constrained \gls{dft}.
(c) Fe moments in the lowest sampled zero-field spiral. Color gives the
rotation angle of the N\'eel vector relative to the first Fe plane.
(d) Period under an applied electric field and at zero field with the polar
mode fixed to the corresponding displacement. At $+20$~mV/\AA{} the
sampled uniform state is lower; no spiral period is assigned.
(e) Zero-field scans under axial strain or shear, each referenced to
the uniform state under the same load.}
\label{fig:bfo-spiral}
\end{figure*}

\begin{figure*}[htbp]
\centering
\includegraphics[width=\linewidth]{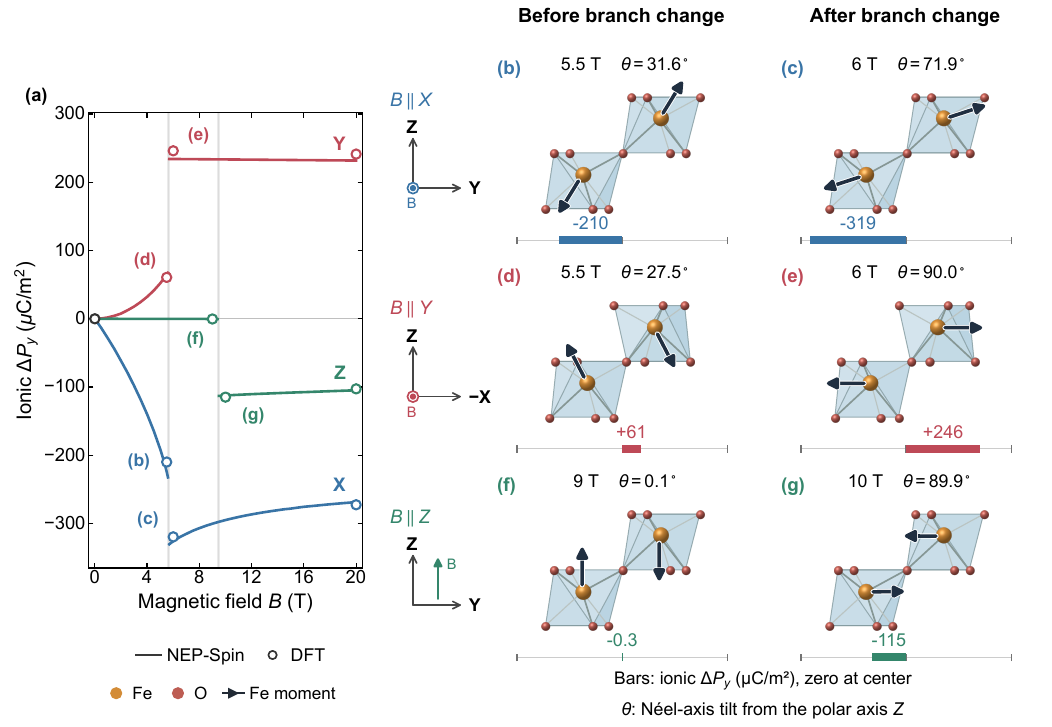}
\caption{Field-induced reorientation of the N\'eel axis and change in
ionic polarization in \gls{bfo}. (a) Lines give $\Delta P_y$ from the
NEP-Spin field scan. Gaps mark a change in the lowest sampled magnetic
branch. Open circles give $\Delta P_y$ from \gls{dft} ionic relaxations
at the corresponding NEP-Spin moment vectors. Each method uses its own
zero-field reference. (b)--(g) Pairs of corner-sharing FeO$_6$ octahedra
from the relaxed \gls{dft} structures before (left) and after (right) each
branch change.
Each pair of panels uses the same view. Arrows show the projected Fe
moments, and $\theta$ is the angle between the N\'eel axis and the polar
axis $Z$. The bars show the change in ionic polarization of the full cell
along $Y$, extending right for positive $\Delta P_y$ and left for negative
$\Delta P_y$. Their colors identify the field directions as in (a).}
\label{fig:bfo-smallcell-field}
\end{figure*}

The BFO fitting set contains 1,658 configurations: 1,355 base configurations, 299 distinct configurations from the property calculations, and four magnetic states along the ferroelectric switching paths. The 1,355 base configurations comprise the 1,133 configurations described in Sec. II C and 222 additional reference configurations. Figure~\ref{fig:bfo-parity}
shows the base set, on which the \gls{rmse} is
0.430~meV/atom for energy, 35.37~meV/\AA{} for atomic forces,
6.62~meV/atom for the virial, and 7.35~meV/$\mu_{\rm B}$
for the Fe magnetic forces. All 459 reference configurations used for the local exchange, anisotropy, volume, and moment scans are included in the fitting set; 207 belong to the base set and 252 lie outside that subset. These scans therefore test energy differences within the fitting set.
Separate checks probe the handedness splitting for two held-out spin configurations
[Fig.~\ref{fig:bfo-spiral}(b)], ionic relaxation at specified Fe moments
(Fig.~\ref{fig:bfo-smallcell-field}), and anisotropy at selected switching-path
geometries (Appendix~\ref{app:bfo-checks}).

Volume scans at fixed moments and moment-amplitude scans at fixed geometry
probe different curvatures of the energy surface (Appendix~\ref{app:bfo-checks}).
Near the reference, the fixed-ion bulk modulus is 153.7~GPa for NEP-Spin and
154.7~GPa for \gls{dft}. The curvature with respect to $m/m_0$ is 90.76 and
88.16~eV per ten-atom cell, respectively. The mixed derivative with respect
to $V/V_0$ and $m/m_0$, estimated from a separate crossed scan, is
$-2.83$~eV per cell for NEP-Spin and $-3.37$~eV per cell for \gls{dft}.

\begin{figure*}[htbp]
\centering
\includegraphics[width=\textwidth]{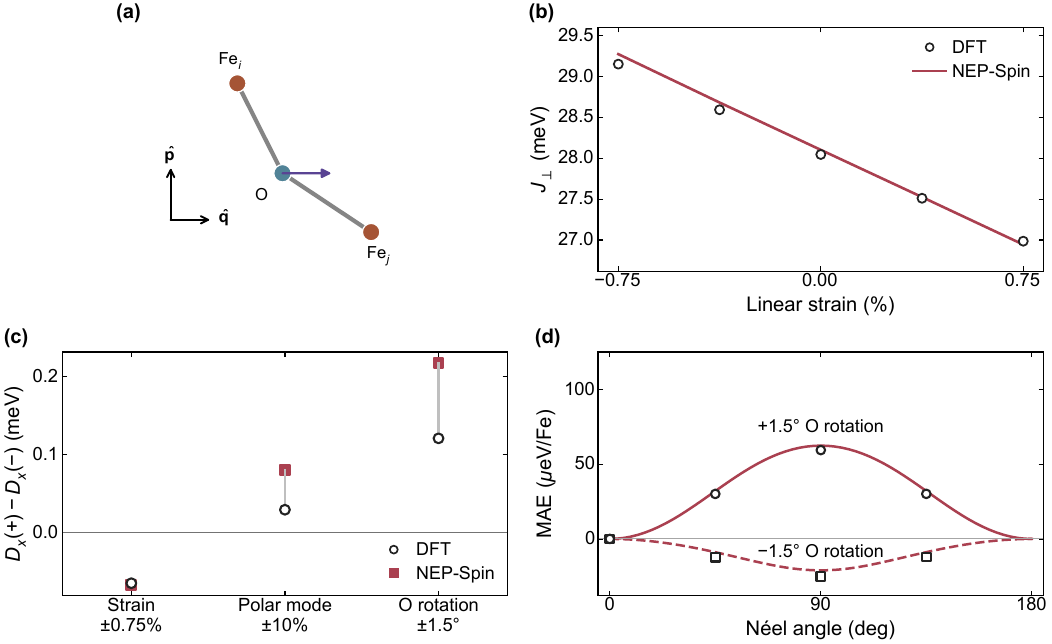}
\caption{Local magnetic interactions in \gls{bfo}.
(a) Projected ordered Fe$_i$--O--Fe$_j$ bridge. The arrow magnifies the
oxygen displacement between the two rotation endpoints twelvefold.
(b) Transverse exchange under isotropic linear strain.
(c) Endpoint difference $\Delta D_x=D_x(+)-D_x(-)$ across
$\pm0.75\%$ linear strain, $\pm10\%$ polar-mode amplitude, and
$\pm1.5^\circ$ oxygen rotation, respectively.
(d) \Glsentrylong{mae} (\glsentryshort{mae}) as the N\'eel axis rotates
from the polar direction $\hat p$ toward the basal direction $\hat q$.
Each angular scan is referenced to its own energy at $0^\circ$.
Open symbols denote \gls{dft} and red marks or curves denote NEP-Spin.}
\label{fig:bfo-magnetic}
\end{figure*}

At fixed Fe--Fe distance, displacing the bridging oxygen atoms changes the
preferred direction of the N\'eel axis [Fig.~\ref{fig:bfo-magnetic}(a),(d)].
The two structures were generated by applying opposite rotational
displacements to the oxygen atoms of the same $R3c$ reference. At a rotation coordinate of
$-1.5^\circ$, \gls{dft} and NEP-Spin favor the basal
direction by 25 and 21~$\mu$eV/Fe, respectively. At $+1.5^\circ$ they
favor the polar direction by 59 and 62~$\mu$eV/Fe.

For the Fe pair in Fig.~\ref{fig:bfo-magnetic}(a), increasing the isotropic
linear strain from $-0.75\%$ to $+0.75\%$ lowers $J_\perp$ by 2.16~meV
in \gls{dft} and 2.33~meV in NEP-Spin. For this bond, NEP-Spin reproduces the sign of the change in $D_x$
for all three distortions [Fig.~\ref{fig:bfo-magnetic}(c)]. The strain
response is $-0.067$~meV, close to the \gls{dft} value of $-0.065$~meV.
The model overestimates the increases induced by polar displacement
($0.080$~meV compared with $0.029$~meV in \gls{dft}) and oxygen rotation
($0.217$~meV compared with $0.121$~meV).

Along two ferroelectric switching paths, the model also reproduces the
easy-axis ordering at the selected intermediate geometries. At one of
these geometries it gives the opposite sign for the shear-pressure
difference (Appendix~\ref{app:bfo-checks}).

The fixed-geometry splitting between opposite spiral handednesses
provides a separate test of chiral interactions. Here $q$ is the signed
wave number along the cycloid propagation axis.
At $|q|=3.73$~nm$^{-1}$ the splitting $E(+q)-E(-q)$ is
$-2.76$~meV/Fe in \gls{dft} and $-2.55$~meV/Fe in NEP-Spin
[Fig.~\ref{fig:bfo-spiral}(b)]. The two spin configurations of opposite chirality at this wave vector were excluded from fitting and used to assess the predicted handedness splitting.

The long-period calculation compares spirals with one turn per cell against
the relaxed uniform \gls{afm} state under the same mechanical constraints.
At zero pressure we relax the ions, Fe moment directions and magnitudes,
and cell degrees of freedom for each sampled branch. A quadratic fit in
$q$ to the lowest spiral energy and its two neighbors gives a period of 69.7~nm
[Fig.~\ref{fig:bfo-spiral}(a)]. The lowest sampled spiral (71.9~nm)
lies 10.2~$\mu$eV/Fe below the lowest sampled uniform state.
The measured bulk period is $62\pm2$~nm~\cite{Sosnowska1982Spiral}, and
a monodomain measurement gives $62.9\pm0.5$~nm at
5~K~\cite{Ramazanoglu2011Period}. The fitted period exceeds the latter
value by 11\%. Within the model, the sampled spiral at 62.94~nm lies
only 0.15~$\mu$eV/Fe above the one at 71.93~nm. The small energy
separation illustrates how weakly the energy selects the period near its minimum.

A longitudinal field acts on the ions
through $\vect F_i^{\rm field}=e(\vect Z_i^*)^{\mathsf T}\vect{\mathcal E}$
using fixed site-resolved Born effective charges. These tensors come
from a separate collinear \gls{pbe}+$U$ linear-response calculation
($U_{\rm eff}=4$~eV, Fe semicore $p$ states in valence, 500~eV cutoff,
$9\times9\times9$ mesh). We subtract the site average of each tensor
component to enforce the acoustic sum rule.
At a given field, spiral and uniform states are compared by their electric
enthalpies. A corresponding zero-field scan fixes the longitudinal polar
displacement to its finite-field value while relaxing the other free
coordinates. We measure the polar-mode amplitude $Q$ by projecting internal
displacements, with homogeneous strain removed, onto the reference
longitudinal polar distortion. Figure~\ref{fig:bfo-spiral}(d) normalizes
this amplitude to its relaxed zero-field value. At $-20$~mV/\AA{} the
amplitude decreases by 5.4\% and the fitted period grows from 69.7 to
93.2~nm. Imposing the same polar displacement at zero field gives
85.1~nm, recovering 15.4~nm of the 23.5-nm increase, or about two-thirds.
At this opposing field the lowest sampled cycloid lies only
1.85~$\mu$eV/Fe below the lowest sampled uniform state in electric enthalpy.
We also applied axial strain along the propagation direction and shear in
the propagation--polar plane, allowing the remaining homogeneous strains
to relax. Across the sampled strains of $\pm0.5\%$, the fitted periods
span 69.1--71.6~nm [Fig.~\ref{fig:bfo-spiral}(e)].

To isolate the ionic polarization response to Fe moment reorientation,
we compare three magnetic field directions in a uniform two-sublattice
\gls{afm} cell [Fig.~\ref{fig:bfo-smallcell-field}].
We take $Z$ along the polar axis, $X$ along the difference of the first
two rhombohedral lattice vectors, and $Y$ along $Z\times X$. At each field
we include the Zeeman energy and relax the ions and both Fe moment vectors
with NEP-Spin at fixed lattice vectors. The lowest sampled magnetic branch
changes between 5.6 and 5.7~T for either in-plane field and between
9.4 and 9.5~T for $B\parallel Z$. Across these intervals the Fe moments
remain nearly antiparallel ($178.6^\circ$--$180^\circ$), with magnitudes
of about $4.16\,\mu_{\rm B}$ at all six points shown in
Fig.~\ref{fig:bfo-smallcell-field}(b)--(g).

For the two in-plane fields, the N\'eel axis tilts away from the polar
axis between 5.5 and 6~T, while the ionic polarization changes in
opposite directions. To test the ionic response to this reorientation,
we hold both Fe moments at their NEP-Spin values and independently relax
all ions with \gls{dft}. At 6~T, the resulting $\Delta P_y$ is $-319$ and
$+246$~$\mu$C/m$^2$ for $B\parallel X$ and $B\parallel Y$, respectively.
NEP-Spin reproduces the opposite signs and the magnitudes, giving
$-326$ and $+234$~$\mu$C/m$^2$. For $B\parallel Z$, the N\'eel axis lies within
$0.1^\circ$ of $Z$ at 9~T and within $0.1^\circ$ of the basal plane at
10~T [Fig.~\ref{fig:bfo-smallcell-field}(f),(g)]. For the high-field
orientation shown, the \gls{dft} ionic response reaches
$-115$~$\mu$C/m$^2$, compared with $-0.3$~$\mu$C/m$^2$ at 9~T.
NEP-Spin gives $-112$ and $-0.3$~$\mu$C/m$^2$ at 10 and 9~T,
respectively. At 20~T the two methods agree to within
10~$\mu$C/m$^2$ for all three field directions.

\subsection{Cycloid across a ferroelectric twin}

We constructed a periodic cell containing two charge-neutral $(110)$ walls
between the pseudocubic $[111]$ (A) and $[11\bar 1]$ (B) polar variants.
The cell spans $301\times86\times1.38$~nm$^3$ and contains 2,787,840
atoms. Both domains start with the same 74.7-nm cycloid in the plane
spanned by the propagation direction and the polar axis of A. At fixed
cell vectors, we relaxed the atomic positions and the directions and
magnitudes of the Fe moments using \gls{fire}. At the limit of 2,500 iterations, the largest atomic
force component was $3.9\times10^{-5}$~eV/\AA{} and the interior polar
displacements differed in direction by $71.7^\circ$.

\begin{figure*}[htbp]
\centering
\includegraphics[width=\linewidth]{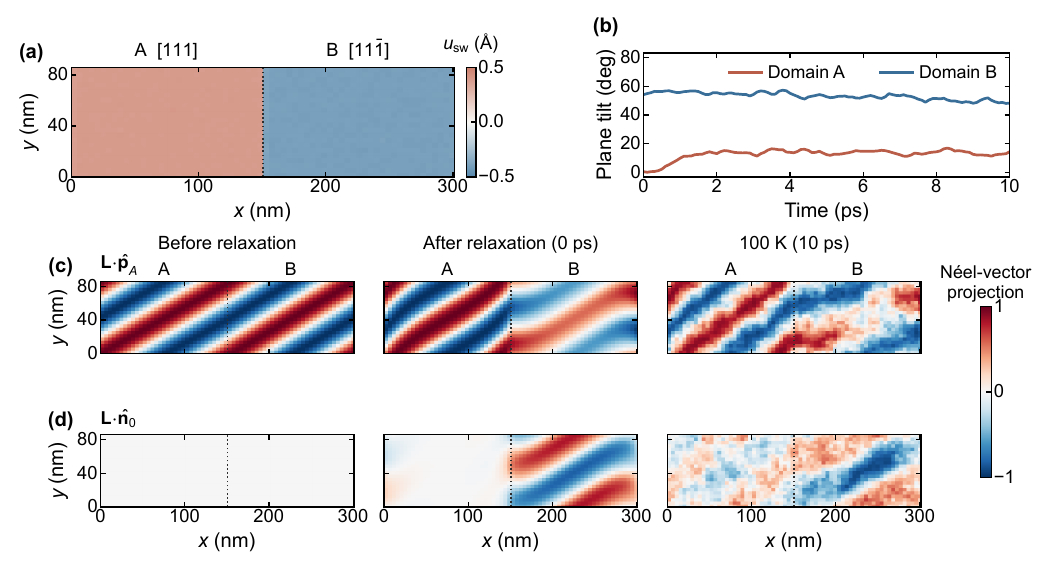}
\caption{Domain dependence of the cycloid spin rotation plane across a ferroelectric twin in zero field.
(a) At 10~ps, the Bi--O displacement projected along the difference of the
two polar unit vectors distinguishes domains A and B. (b) Tilt of the
fitted spin plane in each domain during the 100~K trajectory. The common
plane before relaxation defines zero tilt. The relaxed configuration defines
zero time. (c,d) The same region before relaxation, immediately after
relaxation, and in a single frame at 10~ps, from left to right. Colors give the tile-averaged
staggered Fe unit moments projected onto (c) the polar axis of A,
$\hat{\vect p}_A$, or (d) the normal to the initial spin plane,
$\hat{\vect n}_0$. The stripes in (c) trace the spatial modulation
across both domains. Stripes in (d) identify a component outside the
initial spin plane. At 10~ps these stripes are visible in both domains
and have greater contrast in B. All six magnetic maps use the same color scale.
Dotted lines mark the central twin wall. The second wall lies at the periodic
boundary ($x=0$ and $301$~nm).}
\label{fig:bfo-twin}
\end{figure*}

Starting from the relaxed configuration, we evolved the ions and spins in
zero field at 100~K and analyzed the first 10~ps. We integrated the fixed-length Fe moments
with the semi-implicit B (SIB) scheme~\cite{Mentink2010SIB}, using a
0.5-fs time step and Gilbert damping $\alpha=0.05$.
The two ferroelectric domains are distinguished by the Bi displacement
relative to oxygen [Fig.~\ref{fig:bfo-twin}(a)]. The cycloid stripes in
Fig.~\ref{fig:bfo-twin}(c) retain a similar orientation on both sides of
the wall. To estimate the propagation direction in each domain interior,
we project the staggered Fe moments onto their fitted spin plane and
average the phase increments between neighboring spatial tiles. At 10~ps the
resulting in-plane wave-vector axes differ from their initial directions
by about $7^\circ$ in A and $5^\circ$ in B.

The spin rotation plane changes much more strongly in B.
Before relaxation, the staggered moments in both domains lie in the
same plane. After relaxation, their component normal to this plane shows
pronounced stripes in B and remains small in A [Fig.~\ref{fig:bfo-twin}(d)].
Each tilt in Fig.~\ref{fig:bfo-twin}(b) is the angle between the fitted
spin plane of that domain and the plane shared by A and B before relaxation.
Fitting the staggered moments in the central half of each domain gives
post-relaxation tilts of $0.6^\circ$ in A and $54.1^\circ$ in B.
At 100~K the mean tilts over
5--10~ps are $13.7^\circ$ in A and $51.3^\circ$ in B. The weak stripes
in A at 10~ps [Fig.~\ref{fig:bfo-twin}(d)] reveal a component of the existing cycloid
normal to its initial spin plane. Fitting this projection in the domain
interior to a sinusoid at the initial wave vector gives similar mean
amplitudes over 6--8~ps (0.19) and 8--10~ps (0.20).
For the imposed initial propagation axis,
the plane spanned by that axis and the polar direction of B tilts by
$55.9^\circ$ from the initial plane. The relaxed spin plane in B lies
within $2.2^\circ$ of this geometric reference. A similar relation between
polar and magnetic order was inferred from room-temperature neutron
diffraction by Lebeugle \textit{et al.}~\cite{Lebeugle2008SpinFlop}. After partial $71^\circ$ polarization switching in a BiFeO$_3$ single
crystal, the two domains shared a cycloid propagation vector but had
distinct spin rotation planes.

In these simulations the finite cutoff excludes long-range dipolar fields.
The adiabatic energy supplies instantaneous forces and torques. Simulating
longitudinal dynamics also requires an equation of motion for the moment magnitudes.

\subsection{GPU performance of coupled spin--lattice dynamics}

\begin{figure*}[htbp]
\centering
\includegraphics[width=\linewidth]{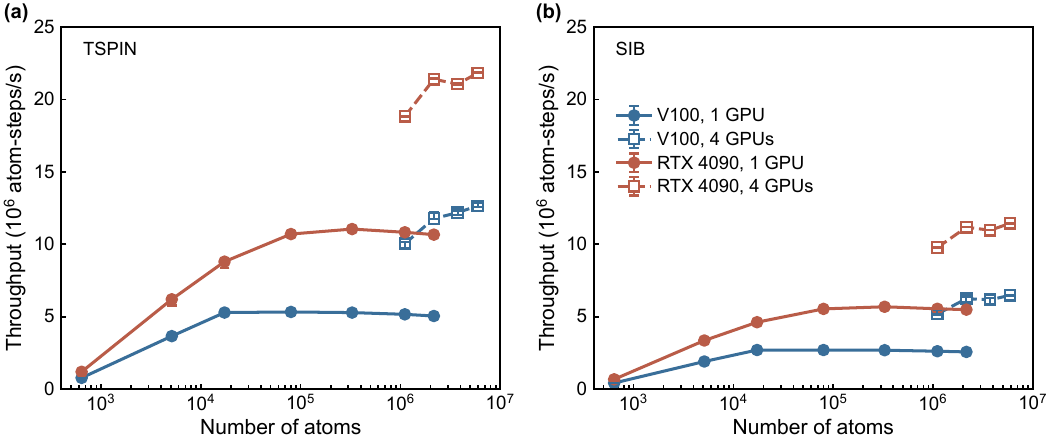}
\caption{NEP-Spin throughput for coupled spin and lattice dynamics in
BiFeO$_3$ using (a) TSPIN and (b) SIB. Circles and squares denote one
and four GPUs, respectively. Both panels use the same scales.
Periodic $R3c$ cells were evolved at 100~K with fixed cell vectors and
a 0.5-fs time step on Tesla V100-SXM2 (32~GB) or GeForce RTX 4090
(24~GB) GPUs. Throughput is the number of atom-steps per second of
integration wall time. Symbols give the median of three consecutive
timing intervals after warm-up. Error bars span their full range.}
\label{fig:md-throughput}
\end{figure*}

Figure~\ref{fig:md-throughput} compares two integrators for the same
\gls{bfo} potential in GPUMD. For a fixed system of 2.16 million atoms,
using four V100 GPUs instead of one increases throughput by factors of
2.33 for the inertial spin integrator (TSPIN) and 2.43 for the
fixed-length semi-implicit integrator (SIB)~\cite{Mentink2010SIB}.
The corresponding speedups on RTX 4090 GPUs are 2.01 and 2.04.

At 5.93 million atoms, four V100 GPUs sustain 12.7 million atom-steps
per second with TSPIN and 6.48 million with SIB. Four RTX 4090 GPUs
reach 21.8 and 11.4 million atom-steps per second, respectively.
The roughly twofold difference between integrators is consistent with
the number of potential evaluations per step. TSPIN evaluates the
potential once. SIB evaluates it at the predicted spin midpoint to
correct the spin directions, then recomputes the atomic and magnetic
forces at the corrected endpoint.

\section{Conclusions}
\label{sec:conclusions}

NEP-Spin constructs a single spin--lattice energy from local bond and
moment tensors. Independent spatial and spin rotations admit exchange
terms that depend on the alignment of neighboring moments. Coupling the
same tensors in a common frame also admits magnetic anisotropy and
antisymmetric exchange. Derivatives of the energy give atomic and magnetic
forces and the configurational virial.

In BiFeO$_3$, displacing the bridging oxygen atoms at fixed Fe--Fe distance
changes the preferred antiferromagnetic axis from basal to polar in
both \gls{dft} and NEP-Spin. The model overestimates the change
in local \gls{dmi} under this distortion. On longer length scales, the lowest
sampled zero-field cycloid lies 10.2~$\mu$eV/Fe below the relaxed uniform
state. Its fitted period of 69.7~nm exceeds the 5~K neutron value by
11\%. An electric field of $-20$~mV/\AA{} reduces the polar-mode amplitude
by 5.4\% and lengthens the period to 93.2~nm. Constraining the same
polar displacement at zero field recovers 15.4~nm of the 23.5-nm increase.

In the uniform antiferromagnetic cell at fixed lattice vectors, a 6~T
field along $X$ decreases the basal-plane ionic polarization component
$P_y$, whereas a field along $Y$ increases it. \gls{dft} ionic
relaxations with the Fe moment vectors fixed to their NEP-Spin values
reproduce both signs.

At a $71^\circ$ ferroelectric twin, the two domains start with a common
cycloid. Its rotation plane tilts primarily in domain B during relaxation.
At 100~K the cycloid in A also develops a component normal to the initial spin plane.
The mean tilts over 5--10~ps are $13.7^\circ$ in A and $51.3^\circ$ in B.
At 10~ps both propagation axes lie within about $7^\circ$ of their
initial directions.
With the same potential, both integrators reach 5.93 million atoms on
four GPUs. At the matched size of 2.16 million atoms, distributing the
calculation over four GPUs roughly halves the time per step on both
hardware platforms. The same BFO potential thus connects changes in local magnetic
interactions under lattice distortions to the evolution of cycloid rotation planes
in a million-atom ferroelectric twin.

\section*{Data and code availability}
The collinear Fe--Al and noncollinear Fe reference datasets are available
through Refs.~\cite{Kotykhov2024Fitting} and \cite{Zheng2026Fe}, respectively.
The trained NEP-Spin models and the data generated in this study are
available from the corresponding author upon reasonable request.
The GPUMD source repository is \url{https://github.com/brucefan1983/GPUMD}.

\begin{acknowledgments}
This work was supported by the Science Challenge Project (No.~TZ2025017).
The authors thank the Open Source Supercomputing Center of S-A-I for providing the NVIDIA V100 GPU computational resources.
\end{acknowledgments}

\section*{Contributions}
Chengbing Chen implemented the software, trained the models, and wrote the
first draft of the manuscript. Zheyong Fan guided the design of the software
architecture and reviewed the code. Ben Xu helped shape the NEP-Spin
framework, supervised the project, and guided its scientific direction.
All authors discussed the results and contributed to revising the manuscript.

\appendix
\section{Radial basis, on-site terms, and descriptor dimensions}
\label{app:grammar}

For a pair cutoff $R=(r^{\rm s}_{Z_i}+r^{\rm s}_{Z_k})/2$, the radial basis
in Eq.~\eqref{eq:compressed-radial} uses the \gls{nep}
Chebyshev expansion~\cite{Xu2025GPUMD} on $0\le r<R$,
\begin{equation}
 f_b(r;R)=\frac{1+T_b(2(r/R-1)^2-1)}{2}
          \frac{1+\cos(\pi r/R)}{2}.
\end{equation}
The functions and their first derivatives vanish at $R$ and the functions
are zero outside. The radial mixing matrix $A$ is shared across species
pairs. Species resolution enters through the ordered pair of indices on
$h^{Z_i Z_k}_{nb}$ in Eq.~\eqref{eq:compressed-radial}.

With the component normalization $Y_{00}=1$ and
$\sum_m Y_{gm}(\hat\rvec)^2=2g+1$, the on-site terms in
Eqs.~\eqref{eq:soc-density} and~\eqref{eq:product-density} are
\begin{align}
 H^{ic,\mathrm{on}}_{Jp\eta,M}
 &=\delta_{p,+1}\sum_{d>0:\,(-1)^d=\eta}
       D_{dJc}\,\Phi_{dJM}(\spinvec_i),\nonumber\\
 T^{ic,\mathrm{on}}_{gj,ma}
 &=\delta_{g,0}\sum_{d>0}D_{djc}\,\Phi_{dja}(\spinvec_i).
\end{align}
Each sum includes only allowed $(d,j)$ pairs of
Eq.~\eqref{eq:spin-primitives}. $D$ is learned independently of the
edge-path weights $B$ in each model. For $g=0$ the spatial index has a
single value. The direct on-site scalar inputs are
$|\spinvec_i|^2,|\spinvec_i|^4,\ldots,|\spinvec_i|^{2\lfloor S/2\rfloor}$.

Let $\mathcal K(L,S)$ be the set of distinct $(J,(-1)^g,(-1)^d)$ reached
by Eq.~\eqref{eq:cg}. The numbers of scalar coordinates appended to the
structural descriptor are
\begin{align}
 d_{\rm SOC}&=\lfloor S/2\rfloor+C\bigl(1+|\mathcal K(L,S)|\bigr),\nonumber\\
 d_{\rm iso}&=\lfloor S/2\rfloor+C\bigl[1+(L+1)(\min(L,S)+1)\bigr].
 \label{eq:channel-count}
\end{align}
The first term in each expression counts local moment powers. Within each channel there is
one signed scalar in addition to the sector norms. For $L=S=2$ there are
ten joint sectors with \gls{soc} and nine product sectors without \gls{soc}. At $C=3$
this gives 34 and 31 coordinates, respectively. $L$ sets angular
resolution, $S$ bounds the degree of the input spin polynomials, and $C$
counts independent radial channels. Products in a norm can reach spin
degree $2S$, and the nonlinear atomic network mixes these scalar
coordinates further. The primitive truncations therefore impose no
corresponding polynomial-degree bound on the fitted energy.

\section{Optional magnetic loss terms}
\label{app:loss-options}

In Eq.~\eqref{eq:mag-loss}, $\mathcal A_\tau$ contains the $N_\tau$ sites with
$|\spinvec_i|^2>10^{-20}\,\mu_{\rm B}^2$.
The torque term weights the squared transverse magnetic-force error by
$|\spinvec_i|^2$. It supplies no information on longitudinal errors.

To weight the two components separately, we define $u_i=\hat{\spinvec}_i\cdot\Delta\mforce_i$ and
$\vect v_i=\Delta\mforce_i-u_i\hat{\spinvec}_i$ for a nonzero moment.
The split objective replaces the Cartesian term by
\begin{equation}
 \mathcal L_M=\lambda_\parallel R_\parallel+\lambda_\perp R_\perp,
 \quad R_\parallel^2=\frac{\sum_i u_i^2}{N_M},\quad
 R_\perp^2=\frac{\sum_i|\vect v_i|^2}{2N_M}.
 \label{eq:split-loss}
\end{equation}
Both sums use $\mathcal A$. At exactly zero moment the two numerators
receive one-third and two-thirds of $|\Delta\mforce_i|^2$, respectively.
Thus $3R_M^2=R_\parallel^2+2R_\perp^2$. The optional torque term can be
used with either choice of magnetic-force supervision.

\section{Additional \texorpdfstring{BiFeO$_3$}{BiFeO3} property comparisons}
\label{app:bfo-checks}

\begin{figure*}[htbp]
\centering
\includegraphics[width=\textwidth]{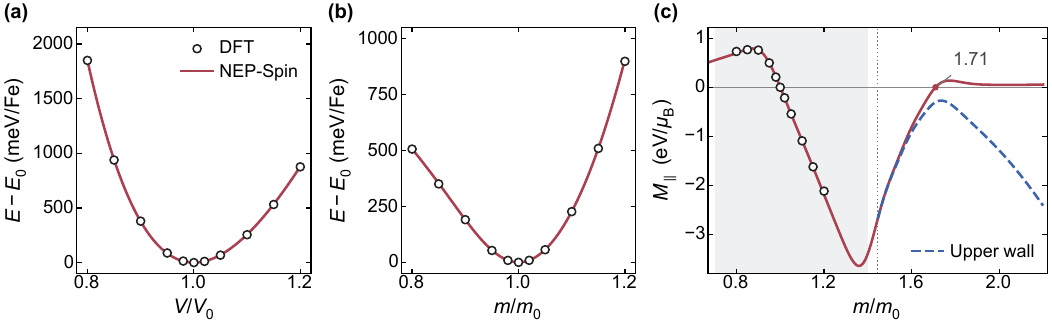}
\caption{Response of \gls{bfo} to volume and Fe moment magnitude.
Energy per Fe relative to the reference at $V_0,m_0$ for (a) a volume scan
at fixed fractional coordinates and Fe moments and (b) common scaling of
the Fe moments at fixed geometry and spin directions.
(c) Mean longitudinal magnetic force per Fe along the extended path in (b),
with positive values pointing toward larger moments. Open circles denote
constrained \gls{dft} data and solid curves denote NEP-Spin.
Shading marks the Fe moment range in the joint training set. The dotted
line marks the illustrative upper threshold $s^{\rm hi}=6.0\,\mu_{\rm B}$,
and the dashed curve includes the wall contribution for
$k^{\rm hi}=0.25\,\mathrm{eV}/\mu_{\rm B}^3$. Here $m_0=4.156\,\mu_{\rm B}$.}
\label{fig:bfo-eos}
\end{figure*}

Figure~\ref{fig:bfo-eos} shows the energy cost of changing the volume at fixed
Fe moments and of changing the Fe moment magnitudes at fixed geometry.
On extending the latter scan, the model's longitudinal magnetic force
changes from inward to outward at $m=1.71m_0$ ($7.10\,\mu_{\rm B}$),
above the largest Fe moment in the training set ($5.81\,\mu_{\rm B}$).
Adding the analytic contribution of the illustrative upper wall keeps the
total force inward from the wall threshold to the scan endpoint at $2.2m_0$.
The constrained \gls{dft} points on this path cover $0.8\le m/m_0\le1.2$.

\begin{figure*}[htbp]
\centering
\includegraphics[width=\textwidth]{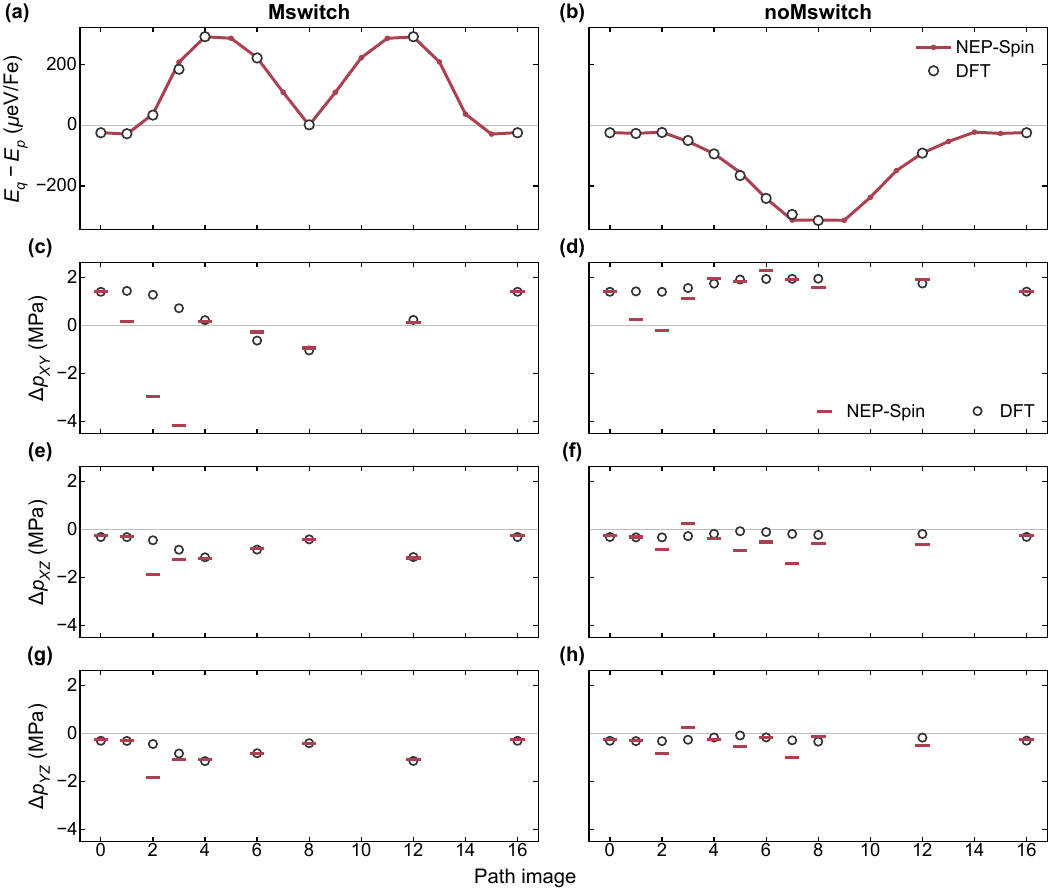}
\caption{Magnetic energy and shear-pressure differences along the two
ferroelectric switching paths of Ref.~\cite{Munro2018Pathways}.
The left and right columns follow the Mswitch and noMswitch geometries,
respectively. At each fixed geometry the G-type antiferromagnetic order
is constrained along the reference polar axis $p$ or the in-plane axis $q$.
(a,b) Energy difference $E_q-E_p$ per Fe.
(c--h) Pressure differences
$\Delta p_{ij}=p_{ij}(q)-p_{ij}(p)$ in a common reference frame
with $Z\parallel p$, $X\parallel q$, and $Y=Z\times X$.
We use the compression-positive convention
$p_{ij}=-\sigma^{\rm sym}_{ij}=W^{\rm sym}_{ij}/V$.
Open circles denote our constrained \gls{dft} results.
The red line connects NEP-Spin energies at all 17 images of each path;
red ticks give NEP-Spin pressures at the images with \gls{dft} references.}
\label{fig:bfo-switch-response}
\end{figure*}

We use the nudged elastic band images supplied by Munro \textit{et al.}~\cite{Munro2018Pathways}
to probe magnetic anisotropy along two ferroelectric switching paths
(Fig.~\ref{fig:bfo-switch-response}). At image 3, \gls{dft} gives
$E_q-E_p=+184$ and $-50$~$\mu$eV/Fe on the Mswitch and noMswitch paths,
respectively. NEP-Spin gives $+208$ and $-53$~$\mu$eV/Fe.
Six path geometries were excluded from fitting, including image 3 on both
paths. Image 7 on the noMswitch path was reserved for a separate assessment.
The shear response is less accurate at the same geometries.
For example, on the Mswitch path $\Delta p_{XY}$ is $+0.72$~MPa
in \gls{dft} and $-4.18$~MPa in NEP-Spin.

\bibliography{references}

@article{Drautz2020Atomic,
  author = {Drautz, Ralf},
  title = {Atomic cluster expansion of scalar, vectorial, and tensorial properties including magnetism and charge transfer},
  journal = {Phys. Rev. B},
  volume = {102},
  pages = {024104},
  year = {2020},
  doi = {10.1103/PhysRevB.102.024104}
}

@article{Rinaldi2024Noncollinear,
  author = {Rinaldi, Matteo and Mrovec, Matous and Bochkarev, Anton and Lysogorskiy, Yury and Drautz, Ralf},
  title = {Non-collinear magnetic atomic cluster expansion for iron},
  journal = {npj Comput. Mater.},
  volume = {10},
  pages = {12},
  year = {2024},
  doi = {10.1038/s41524-024-01196-8}
}

@article{Ma2008Largescale,
  author = {Ma, Pui-Wai and Woo, C. H. and Dudarev, S. L.},
  title = {Large-scale simulation of the spin-lattice dynamics in ferromagnetic iron},
  journal = {Phys. Rev. B},
  volume = {78},
  pages = {024434},
  year = {2008},
  doi = {10.1103/PhysRevB.78.024434}
}

@article{Fan2021Neuroevolution,
  author = {Fan, Zheyong and Zeng, Zezhu and Zhang, Cunzhi and Wang, Yanzhou and Song, Keke and Dong, Haikuan and Chen, Yue and Ala-Nissila, Tapio},
  title = {Neuroevolution machine learning potentials: Combining high accuracy and low cost in atomistic simulations and application to heat transport},
  journal = {Phys. Rev. B},
  volume = {104},
  pages = {104309},
  year = {2021},
  doi = {10.1103/PhysRevB.104.104309}
}

@inproceedings{Schaul2011SNES,
  author = {Schaul, Tom and Glasmachers, Tobias and Schmidhuber, J{\"u}rgen},
  title = {High dimensions and heavy tails for natural evolution strategies},
  booktitle = {Proceedings of the 13th Annual Conference on Genetic and Evolutionary Computation},
  publisher = {ACM}, year = {2011}, pages = {845--852},
  doi = {10.1145/2001576.2001692}
}

@article{Xu2025GPUMD,
  author = {Xu, Ke and Bu, Hekai and Pan, Shuning and Lindgren, Eric and Wu, Yongchao and
    Wang, Yong and Liu, Jiahui and Song, Keke and Xu, Bin and Li, Yifan and
    Hainer, Tobias and Svensson, Lucas and Wiktor, Julia and Zhao, Rui and Huang, Hongfu and
    Qian, Cheng and Zhang, Shuo and Zeng, Zezhu and Zhang, Bohan and Tang, Benrui and
    Xiao, Yang and Yan, Zihan and Shi, Jiuyang and Liang, Zhixin and Wang, Junjie and
    Liang, Ting and Cao, Shuo and Wang, Yanzhou and Ying, Penghua and Xu, Nan and
    Chen, Chengbing and Zhang, Yuwen and Chen, Zherui and Wu, Xin and Jiang, Wenwu and
    Berger, Esme and Li, Yanlong and Chen, Shunda and Gabourie, Alexander J. and Dong, Haikuan and
    Xiong, Shiyun and Wei, Ning and Chen, Yue and Xu, Jianbin and Ding, Feng and
    Sun, Zhimei and Ala-Nissila, Tapio and Harju, Ari and Zheng, Jincheng and Guan, Pengfei and
    Erhart, Paul and Sun, Jian and Ouyang, Wengen and Su, Yanjing and Fan, Zheyong},
  title = {{GPUMD 4.0}: A high-performance molecular dynamics package for versatile materials simulations with machine-learned potentials},
  journal = {Materials Genome Engineering Advances},
  volume = {3},
  number = {3},
  pages = {e70028},
  year = {2025},
  doi = {10.1002/mgea.70028}
}

@article{Eckhoff2021Highdimensional,
  author = {Eckhoff, Marco and Behler, J{\"o}rg},
  title = {High-dimensional neural network potentials for magnetic systems using spin-dependent atom-centered symmetry functions},
  journal = {npj Comput. Mater.},
  volume = {7},
  pages = {170},
  year = {2021},
  doi = {10.1038/s41524-021-00636-z}
}

@article{Novikov2022Magnetic,
  author = {Novikov, Ivan and Grabowski, Blazej and K{\"o}rmann, Fritz and Shapeev, Alexander},
  title = {Magnetic Moment Tensor Potentials for collinear spin-polarized materials reproduce different magnetic states of bcc {Fe}},
  journal = {npj Comput. Mater.},
  volume = {8},
  pages = {13},
  year = {2022},
  doi = {10.1038/s41524-022-00696-9}
}

@article{Domina2022Spectral,
  author = {Domina, M. and Cobelli, M. and Sanvito, S.},
  title = {Spectral neighbor representation for vector fields: Machine learning potentials including spin},
  journal = {Phys. Rev. B},
  volume = {105},
  pages = {214439},
  year = {2022},
  doi = {10.1103/PhysRevB.105.214439}
}

@article{Chapman2022Machinelearned,
  author = {Chapman, Jacob B. J. and Ma, Pui-Wai},
  title = {A machine-learned spin-lattice potential for dynamic simulations of defective magnetic iron},
  journal = {Sci. Rep.},
  volume = {12},
  pages = {22451},
  year = {2022},
  doi = {10.1038/s41598-022-25682-5}
}

@article{Cai2023DeltaSpin,
  author = {Cai, Zefeng and Wang, Ke and Xu, Yong and Wei, Su-Huai and Xu, Ben},
  title = {A self-adaptive first-principles approach for magnetic excited states},
  journal = {Quantum Front.},
  volume = {2},
  number = {1},
  pages = {21},
  year = {2023},
  doi = {10.1007/s44214-023-00041-1}
}

@article{Yu2024Spindependent,
  author = {Yu, Hongyu and Zhong, Yang and Hong, Liangliang and Xu, Changsong and Ren, Wei and Gong, Xingao and Xiang, Hongjun},
  title = {Spin-dependent graph neural network potential for magnetic materials},
  journal = {Phys. Rev. B},
  volume = {109},
  pages = {144426},
  year = {2024},
  doi = {10.1103/PhysRevB.109.144426}
}

@article{Yu2024Physicsinformed,
  author = {Yu, Hongyu and Liu, Boyu and Zhong, Yang and Hong, Liangliang and Ji, Junyi and Xu, Changsong and Gong, Xingao and Xiang, Hongjun},
  title = {Physics-informed time-reversal equivariant neural network potential for magnetic materials},
  journal = {Phys. Rev. B},
  volume = {110},
  pages = {104427},
  year = {2024},
  doi = {10.1103/PhysRevB.110.104427}
}

@article{Kotykhov2023Constrained,
  author = {Kotykhov, Alexey S. and Gubaev, Konstantin and Hodapp, Max and Tantardini, Christian and Shapeev, Alexander V. and Novikov, Ivan S.},
  title = {Constrained {DFT}-based magnetic machine-learning potentials for magnetic alloys: a case study of {Fe--Al}},
  journal = {Sci. Rep.}, volume = {13}, pages = {19728}, year = {2023},
  doi = {10.1038/s41598-023-46951-x}
}

@article{Kotykhov2024Fitting,
  author = {Kotykhov, Alexey S. and Gubaev, Konstantin and Sotskov, Vadim and Tantardini, Christian and Hodapp, Max and Shapeev, Alexander V. and Novikov, Ivan S.},
  title = {Fitting to magnetic forces improves the reliability of magnetic Moment Tensor Potentials},
  journal = {Comput. Mater. Sci.},
  volume = {245}, pages = {113331}, year = {2024},
  doi = {10.1016/j.commatsci.2024.113331}
}

@misc{Ho2026Equivariant,
  author = {Ho, Cheuk Hin and van der Oord, Cas and Darby, James P. and Keane, Theo and Benson, Raz L. and Espinoza, Cristian Rebolledo and Kulkarni, Rutvij and Spinu, Elina and Papanikolaou, Michail and Tomsett, Richard and Forrest, Robert M. and Bean, Jonathan J. and Cs{\'a}nyi, G{\'a}bor and Ortner, Christoph},
  title = {Equivariant Many-body Message Passing Interatomic Potentials for Magnetic Materials},
  year = {2026}, eprint = {2604.08143}, archivePrefix = {arXiv},
  doi = {10.48550/arXiv.2604.08143}
}

@article{Zheng2026Fe,
  author = {Zheng, Daye and Peng, Xingliang and Huang, Yike and Wang, Yinan and Zhang, Duo and Huang, Zhengtao and Cai, Zefeng and Zhang, Linfeng and Chen, Mohan and Xu, Ben and Zhou, Weiqing},
  title = {Integrating deep-learning-based magnetic model and non-collinear spin-constrained method: methodology, implementation and application},
  journal = {npj Comput. Mater.},
  volume = {12}, pages = {52}, year = {2026},
  doi = {10.1038/s41524-025-01923-9}
}

@misc{Gao2026STEP,
  author = {Gao, Yuanqing and Luo, Wen-Hao and Zhang, Lei and Cao, Kun},
  title = {{STEP}: Spin Tensor Equivariant Potential for Data-Efficient Learning of Magnetic Potential Energy Surfaces},
  year = {2026}, eprint = {2607.17129}, archivePrefix = {arXiv},
  doi = {10.48550/arXiv.2607.17129}
}

@article{Yang2024Deep,
  author = {Yang, Teng and Cai, Zefeng and Huang, Zhengtao and Tang, Wenlong and Shi, Ruosong and Godfrey, Andy and Liu, Hanxing and Lin, Yuanhua and Nan, Ce-Wen and Ye, Meng and Zhang, LinFeng and Wang, Ke and Wang, Han and Xu, Ben},
  title = {Deep learning illuminates spin and lattice interaction in magnetic materials},
  journal = {Phys. Rev. B},
  volume = {110}, pages = {064427}, year = {2024},
  doi = {10.1103/PhysRevB.110.064427}
}

@article{Xu2019Magnetic,
  author = {Xu, Changsong and Xu, Bin and Dup{\'e}, Bertrand and Bellaiche, L.},
  title = {Magnetic interactions in {BiFeO$_3$}: A first-principles study},
  journal = {Phys. Rev. B},
  volume = {99}, pages = {104420}, year = {2019},
  doi = {10.1103/PhysRevB.99.104420}
}

@article{Streib2022Adiabatic,
  author = {Streib, Simon and Cardias, Ramon and Pereiro, Manuel and Bergman, Anders and Sj{\"o}qvist, Erik and Barreteau, Cyrille and Delin, Anna and Eriksson, Olle and Thonig, Danny},
  title = {Adiabatic spin dynamics and effective exchange interactions from constrained tight-binding electronic structure theory: Beyond the {Heisenberg} regime},
  journal = {Phys. Rev. B},
  volume = {105}, pages = {224408}, year = {2022},
  doi = {10.1103/PhysRevB.105.224408}
}

@article{Yuan2024MagNet,
  author = {Yuan, Zilong and Xu, Zhiming and Li, He and Cheng, Xinle and Tao, Honggeng and Tang, Zechen and Zhou, Zhiyuan and Duan, Wenhui and Xu, Yong},
  title = {Equivariant neural network force fields for magnetic materials},
  journal = {Quantum Front.},
  volume = {3}, pages = {8}, year = {2024},
  doi = {10.1007/s44214-024-00055-3}
}

@misc{Wang2026HotPPSpin,
  author = {Wang, Junjie and Zhu, Yijie and Zhang, Zhongwei and Guo, Zhiyue and Zhu, Xudong and He, Lixin and Ding, Chi and Sun, Jian},
  title = {Cartesian tensor equivariant machine-learning force field for spin-dependent atomistic simulations},
  year = {2026}, eprint = {2608.30338}, archivePrefix = {arXiv},
  primaryClass = {physics.comp-ph}, doi = {10.48550/arXiv.2608.30338}
}

@article{Kresse1994AbInitio,
  title = {{\emph{Ab Initio}} Molecular-Dynamics Simulation of the Liquid-Metal--Amorphous-Semiconductor Transition in Germanium},
  author = {Kresse, G. and Hafner, J.},
  year = {1994},
  month = may,
  journal = {Phys. Rev. B},
  volume = {49},
  number = {20},
  pages = {14251--14269},
  publisher = {American Physical Society},
  issn = {0163-1829, 1095-3795},
  doi = {10.1103/PhysRevB.49.14251},
  urldate = {2024-03-27}
}

@article{Kresse1996Efficient,
  title = {Efficient Iterative Schemes for {\emph{Ab Initio}} Total-Energy Calculations Using a Plane-Wave Basis Set},
  author = {Kresse, G. and Furthm{\"u}ller, J.},
  year = {1996},
  month = oct,
  journal = {Phys. Rev. B},
  volume = {54},
  number = {16},
  pages = {11169--11186},
  publisher = {American Physical Society},
  issn = {0163-1829, 1095-3795},
  doi = {10.1103/PhysRevB.54.11169},
  urldate = {2024-03-27}
}

@article{Blochl1994Projector,
  title = {Projector Augmented-Wave Method},
  author = {Bl{\"o}chl, P. E.},
  year = {1994},
  month = dec,
  journal = {Phys. Rev. B},
  volume = {50},
  number = {24},
  pages = {17953--17979},
  publisher = {American Physical Society},
  issn = {0163-1829, 1095-3795},
  doi = {10.1103/PhysRevB.50.17953},
  urldate = {2024-03-27}
}

@article{Perdew1996Generalized,
  title = {Generalized {{Gradient Approximation Made Simple}}},
  author = {Perdew, John P. and Burke, Kieron and Ernzerhof, Matthias},
  year = {1996},
  month = oct,
  journal = {Phys. Rev. Lett.},
  volume = {77},
  number = {18},
  pages = {3865--3868},
  publisher = {American Physical Society},
  issn = {0031-9007, 1079-7114},
  doi = {10.1103/PhysRevLett.77.3865},
  urldate = {2024-03-27}
}

@article{Chen2025NepTrain,
  author = {Chen, Chengbing and Li, Yutong and Zhao, Rui and Liu, Zhoulin and Fan, Zheyong and Tang, Gang and Wang, Zhiyong},
  title = {{NepTrain} and {NepTrainKit}: Automated active learning and visualization toolkit for neuroevolution potentials},
  journal = {Comput. Phys. Commun.},
  volume = {317},
  pages = {109859},
  year = {2025},
  doi = {10.1016/j.cpc.2025.109859}
}

@article{Kresse1999PAW,
  author = {Kresse, G. and Joubert, D.},
  title = {From ultrasoft pseudopotentials to the projector augmented-wave method},
  journal = {Phys. Rev. B},
  volume = {59},
  pages = {1758--1775},
  year = {1999},
  doi = {10.1103/PhysRevB.59.1758}
}

@article{Dudarev1998LSDAU,
  author = {Dudarev, S. L. and Botton, G. A. and Savrasov, S. Y. and Humphreys, C. J. and Sutton, A. P.},
  title = {Electron-energy-loss spectra and the structural stability of nickel oxide: An {LSDA+U} study},
  journal = {Phys. Rev. B},
  volume = {57},
  pages = {1505--1509},
  year = {1998},
  doi = {10.1103/PhysRevB.57.1505}
}

@article{Sosnowska1982Spiral,
  author = {Sosnowska, I. and Peterlin-Neumaier, T. and Steichele, E.},
  title = {Spiral magnetic ordering in bismuth ferrite},
  journal = {J. Phys. C: Solid State Phys.},
  volume = {15},
  pages = {4835--4846},
  year = {1982},
  doi = {10.1088/0022-3719/15/23/020}
}

@article{Ramazanoglu2011Period,
  author = {Ramazanoglu, M. and Ratcliff, II, W. and Choi, Y. J. and Lee, Seongsu and Cheong, S.-W. and Kiryukhin, V.},
  title = {Temperature-dependent properties of the magnetic order in single-crystal {BiFeO$_3$}},
  journal = {Phys. Rev. B},
  volume = {83},
  pages = {174434},
  year = {2011},
  doi = {10.1103/PhysRevB.83.174434}
}

@article{Lebeugle2008SpinFlop,
  author = {Lebeugle, D. and Colson, D. and Forget, A. and Viret, M. and Bataille, A. M. and Gukasov, A.},
  title = {Electric-Field-Induced Spin Flop in {BiFeO$_3$} Single Crystals at Room Temperature},
  journal = {Phys. Rev. Lett.},
  volume = {100},
  pages = {227602},
  year = {2008},
  doi = {10.1103/PhysRevLett.100.227602}
}

@article{Mentink2010SIB,
  author = {Mentink, J. H. and Tretyakov, M. V. and Fasolino, A. and Katsnelson, M. I. and Rasing, Th.},
  title = {Stable and fast semi-implicit integration of the stochastic {Landau--Lifshitz} equation},
  journal = {J. Phys.: Condens. Matter},
  volume = {22},
  pages = {176001},
  year = {2010},
  doi = {10.1088/0953-8984/22/17/176001}
}

@article{Munro2018Pathways,
  author = {Munro, Jason M. and Akamatsu, Hirofumi and Padmanabhan, Haricharan and Liu, Vincent S. and Shi, Yin and Chen, Long-Qing and VanLeeuwen, Brian K. and Dabo, Ismaila and Gopalan, Venkatraman},
  title = {Discovering minimum energy pathways via distortion symmetry groups},
  journal = {Phys. Rev. B},
  volume = {98},
  pages = {085107},
  year = {2018},
  doi = {10.1103/PhysRevB.98.085107}
}

\end{document}